\documentclass[a4paper,fleqn,usenatbib]{mnras}

\usepackage{mathptmx}
\usepackage[T1]{fontenc}
\usepackage{ae,aecompl}

\usepackage{graphicx}	% Including figure files
\usepackage{amsmath}	% Advanced maths commands
\usepackage{amssymb}	% Extra maths symbols
\usepackage{longtable}
\usepackage{multirow}
\usepackage{enumerate}   
\usepackage{lastpage}
\usepackage{subfig}
\usepackage{lipsum}
\title[Segregation Phenomena and VDPs]{The Influence of the Dynamic State of Galaxy Clusters on Segregation Phenomena and Velocity Dispersion Profiles}

\author[Nascimento, Lopes, Ribeiro, Costa \& Morell]{
R.S. Nascimento,$^{1,2}$\thanks{E-mail: rnascimento@lna.br}
P.A.A. Lopes,$^{2}$
A.L.B. Ribeiro,$^{3}$
A.P. Costa,$^{3}$
and D.F. Morell$^{3,4}$
\\
$^{1}$Laborat\'orio Nacional de Astrof\'isica/MCTI, Itajub\'a 37504-364, Brazil\\
$^{2}$Observat\'orio do Valongo, Universidade Federal do Rio de Janeiro, RJ 20080-090, Brazil\\
$^{3}$Laborat\'orio de Astrof\'isica Te\'orica e Observacional, Universidade Estadual de Santa Cruz, Ilh\'eus, BA 454650-000, Brazil\\
$^{4}$Observat\'orio Nacional, Rio de Janeiro, RJ, Brazil
}

\date{Accepted 2018 November 29. Received 2018 November 29; in original form 2018 August 27}

\pubyear{2017}

\begin{document}
\label{firstpage}
\pagerange{\pageref{firstpage}--\pageref{lastpage}}
\maketitle

\begin{abstract}

In this work we investigate the influence of the dynamic state of galaxy clusters on segregation effects and velocity dispersion profiles (VDPs) for a sample of 111 clusters extracted from SDSS-DR7. We find that 73 clusters have Gaussian (G) velocity distribution and 38 clusters have a complex or non-Gaussian (NG) velocity distribution.  We also split the G and NG samples into 'active' and 'passive' galaxies, according to their sSFRs and stellar masses. Our results indicate a strong spatial segregation between active and passive galaxies both in G and NG systems, with passive galaxies being more central. We also found that the passive population in G systems is the only family with lower velocity dispersions for the brightest galaxies ($M_r \lesssim -22.75$), thus presenting velocity segregation with luminosity. The similarity found between the VDPs of the galaxy populations in NG systems indicate that these sets probably share a similar mix of orbits. We also found a clear evolutionary trend for G systems, with brighter galaxies in richer clusters having flatter VDPs. The scenario emerging from this study suggests a direct relationship between segregation effects, VDPs and the dynamic state of clusters.
\end{abstract}

\begin{keywords}
galaxy clusters -- galaxies -- environmental effects
\end{keywords}

%%%%%%%%%%%%%%%%%%%%%%%%%%%%%%%%%%%%%%%%%%%%%%%%%%

%%%%%%%%%%%%%%%%% BODY OF PAPER %%%%%%%%%%%%%%%%%%

\section{INTRODUCTION}

The study of segregation phenomena may help us understand the interplay between galaxy evolution and cluster environment. Segregation means different galaxy types presenting unequal distributions of their luminosities, clustercentric distances, velocities, and/or colours. A number of such effects have been reported over the years \citep[e.g.][]{MD, Scs89, zbf, Ada98, la, RLM10, bar18}. A common aspect in many of these works is the existence of some kind of kinematic segregation regarding the behaviour of the velocity dispersion for galaxy populations: blue/emission-line/late type galaxies have larger velocity dispersion than red/no emission-line/early type galaxies at high confidence levels \citep[see e.g.][]{go1}. This kinematic segregation with colour can be accompanied by a luminosity segregation effect, in the sense that brighter objects have lower velocities \citep[e.g.][]{Bi92, Gi03}, although only the very luminous galaxies seem to be significantly segregated from the remaining cluster members \citep[see][]{Bi92}. 

Interestingly, several studies have also pointed important differences between galaxies residing in systems classified as Gaussian (G) or non-Gaussian (NG), according to their velocity distributions, which is also a type of segregation. For instance, \citet{H09} investigated the dynamical state of CNOC2 galaxy groups and found that the velocity dispersion profiles of G groups are distinct from those of NG groups. \citet{RLM10} analysed a group sample from 2dF Percolation-Inferred Galaxy Group (2PIGG) and found that galaxies residing in G groups are more evolved than in non-Gaussian systems out to distances of $\sim 4R_{200}$, presenting significantly redder colours. \citet{Mz12} found that G and NG systems have different luminosity functions with the former having a brighter characteristic absolute magnitude and a steeper faint-end slope. They concluded that the dynamical state of massive and luminous groups is directly related to the luminosity of its galaxy members. \citet{H12} analysed 15 rich groups at intermediate redshifts ($z \sim 0.4$) extracted from Group Environmental and Evolution Collaboration catalogue and found that the colour distribution of systems with substructure (most of them being NG systems) differ significantly from those without substructure (some of them considered as G systems). In addition, they pointed that groups with substructure show a dominant blue and active star-forming galaxy populations.  Also, \citet{R13}  studied a sample from the SDSS group catalogue identified by \citet{Ber06} and found that there is a significant difference between
the galaxy properties of the inner and outer galaxy populations, suggesting that the environment is actively affecting the galaxies. \citet{deC17} examined a sample of rich clusters from the SDSS Yang catalogue \citep{y07} and found that faint galaxies in NG groups are mainly infalling for the first time into the clusters. Finally, \citet{rob18} showed that the median  X-ray asymmetry  of NG clusters is larger than that of G clusters. All these findings suggest the intrincate relationship between galaxy evolution and the cluster assembly history. 

In \citet{NRL} we studied segregation effects in groups at higher redshifts ($z \sim 0.6$) but for these systems was not possible to split the sample according to the dynamic state of the groups. Now, we investigate the influence of the dynamic state of low-{\it z} galaxy clusters on segregation effects and velocity dispersion profiles. In Section 2, we present our data and stacked samples. The analysis is presented in Section 3. In section 3 we give a summary of the results and our conclusions. Throughout this paper we adopt the cosmology with $\Omega_m = 0.3$,  $\Omega_\Lambda = 0.7$ and H$_0$ = 100 {\it h} km$^{-1}$ s$^{-1}$ Mpc$^{-1}$, with {\it h} set to 0.7. 
\vspace{-0.7cm}

\section{The SDSS Group Sample}

Our initial sample contains 10,124 systems and corresponds to an updated version of SDSS group catalogue identified by \citet{Ber06}. This new version is described in \citet{Lab10} and differ from the first one only in the area used (9380 square degrees from DR7, compared to the original area of 3495 square degrees from the DR3). We derived a refined central redshift and member list for each group. The new central redshift is obtained applying the gap technique \citep{Ada98, Lop07, Lop09} to the central (0.67 Mpc) galaxies. The update list of group members is derived through the application of the ``shifting gapper" technique \citep{Lop09}. This method consists of applying the gap-technique in radial bins from the group centre. The bin size is 0.56 Mpc or larger to force the selection of at least 15 galaxies. Galaxies not linked to main body of the group are eliminated. 
The groups are then subject to a virial analysis ana\-lo\-gous to that described in \citet{Gi98,Pop05,Pop07,Bi06,Lop09}. 
%This procedure yields estimates of velocity dispersion ($\sigma_v$), radius ($R_{500}$, $R_{200}$) and mass ($M_{500}$, $M_{200}$) for most of the groups from the FoF sample. 
Our shift gapper $+$ virial analysis code has been compared to a set of 24 galaxy-based cluster mass estimation techniques and proved
to be among the best three \citep{old15}. From this analysis we find that the error associated to
our mass estimate, $M_{200}$, is $\sim$0.22 dex. See details of the code in \citet{Lop09}.

The final sample contains 5,352 groups from the updated FoF group catalogue, with redshift $z_{max}$ = 0.106 and $N_{min}$ = 5. Then, we selected a subsample of 111 groups containing at least 15 galaxies brighter than ${\text M_r} = -20.5$ in $R_{200}$.  Clustercentric radii were computed using the redshift and angular separation between the galaxy
positions and the luminosity-weighted centre of the clusters. The sample has member galaxies with 
stellar mass $10^{9.5} \leq (M_\star/M_\odot) \leq 10^{12}$, being 95\% complete down to the $M_\star$ lower limit.
These systems have velocity dispersions and $M_{200}$ in the ranges $325~\text{kms}^{-1} \leq \sigma_v \leq 950~\text{kms}^{-1}$,  and $1.06 \times 10^{14} \text{M}_\odot \leq M_{200} \leq 1.86 \times 10^{15} \text{M}_\odot$, respectively. The subsequent analysis will be done using this subsample of rich clusters.

Absolute galaxy magnitudes in {\it gr} bands are derived using the formula: $M_{g,r} = m_{g,r} - DM -k_{corr} - Qz$, where {\it DM} is the distance modulus, $k_{corr}$ is the {\it k}-correction and $Q = -1.4$ \citep[][]{YL99} is a mild evolutionary correction applied to the magnitudes. The {\it k}-corrections for the galaxies are obtained directly from the SDSS-DR12 database \citep{Bk16}. Stellar masses ($M_\star$), star formation rates (SFR) and specific star formation rates (sSFR) for the galaxies in our sample have been extracted from the ``Galspec" analysis provided by the MPA-JHU group (Max Planck Institute for Astrophysics and Johns Hopkins University).
%\footnote{http://wwwmpa.mpa-garching.mpg.de/SDSS/DR7}. 
The total stellar masses are based on model magnitudes. Star formation rates are computed within the galaxy fibre aperture using the nebular emission lines as described in \citet{B04}. Outside of the fibre the estimates use the galaxy photometry following \citet{S07}. AGN and galaxies with weak emission lines, have SFRs estimates from the photometry.
\vspace{-0.4cm}

\subsection{Defining the stacked clusters}

The velocity distribution of galaxies in groups and clusters can provide clues about their dynamic state. In general, evolved or relaxed systems are supposed to have Gaussian velocity distributions while those with non-Gaussian distributions are considered as less evolved or unrelaxed systems \citep{YV77,MF96}. Several statistical methods are used to distinguish between relaxed and unrelaxed systems
according to their velocity distributions \citep[e.g.][]{H09,RLM11,R13,deC17}. We use here the Hellinger Distance \citep[HD,][]{Am85} for detecting departures from a Gaussian velocity distribution. The HD is an estimator of the distance between the empirical velocity distribution of galaxies in a cluster and the theoretically expected Gaussian distribution function \citep[see][for a detailed description of this method]{R13,deC17}. We estimate HD using codes available in the R environment under the distrEx package \citep{ruck}. The method is based on a set of 1,000 realizations of the observed sample, and on a comparison of each resample with a Gaussian template calibrated by the sample size. For each sample size a critical HD value allows the classification of a system as G or NG \citep[see][for details]{deC17}.
The recurrence of the diagnostic for a minimum number of times provides the reliability of the result. \citet{deC17} define this minimum as 70\% of the total number of realizations. In the present work, we only consider  Gaussian  (G)  or  Non-Gaussian  (NG)  systems classified with  reliability  greater  than  95\%.

After the classification of clusters using the HD method, we find 73 G systems (4,693 galaxies) and 38 NG systems (2,899 galaxies). Before proceeding to the segregation analysis, we define two stacked clusters, Gaussian - G  and non-Gaussian - NG. Galaxies in these composite clusters have distances to cluster centres normalized by $R_{200}$ and their velocities are referred to the cluster median velocities and scaled by the cluster velocity dispersion. The normalized velocity dispersion of the combined system, $\sigma_u$, is related to the dimensionless quantity $u_i = \frac{v_i - \langle v \rangle_j}{\sigma_j}$,
where $i$ and $j$ are the galaxy and clusters index, respectively. 
We also define sub-samples from these two initial stacked systems according to the evolutionary stage of galaxies. We separate 'more evolved' from 'less evolved' galaxies based on their sSFRs and stellar masses (${\rm M_\ast}$), using the line defined by \citet{OH16}: 

\vspace{-0.5cm}
\begin{equation}
{\rm log_{10}(sSFR_{cut}/yr^{-1})} = -0.4*{\rm log_{10}(M_\ast / M_{\sun})} - 6.6
\label{eq2}
\end{equation}

\noindent which allow us to split the G and NG samples into star forming or 'active' galaxies and 'passive' ones. Finally, we have four composite systems: PG (passive galaxies in G clusters), AG (active galaxies in G clusters), PNG (passive galaxies in NG clusters), and ANG (active galaxies in NG clusters). These samples would be less than 4\% different if we had used a division at 
${\rm log_{10}(sSFR/yr^{-1})} = -11$, as indicated by \citet{wtc12}.

\begin{figure}
  \centering
  \begin{tabular}{@{}c@{}}
    \includegraphics[scale=0.45]{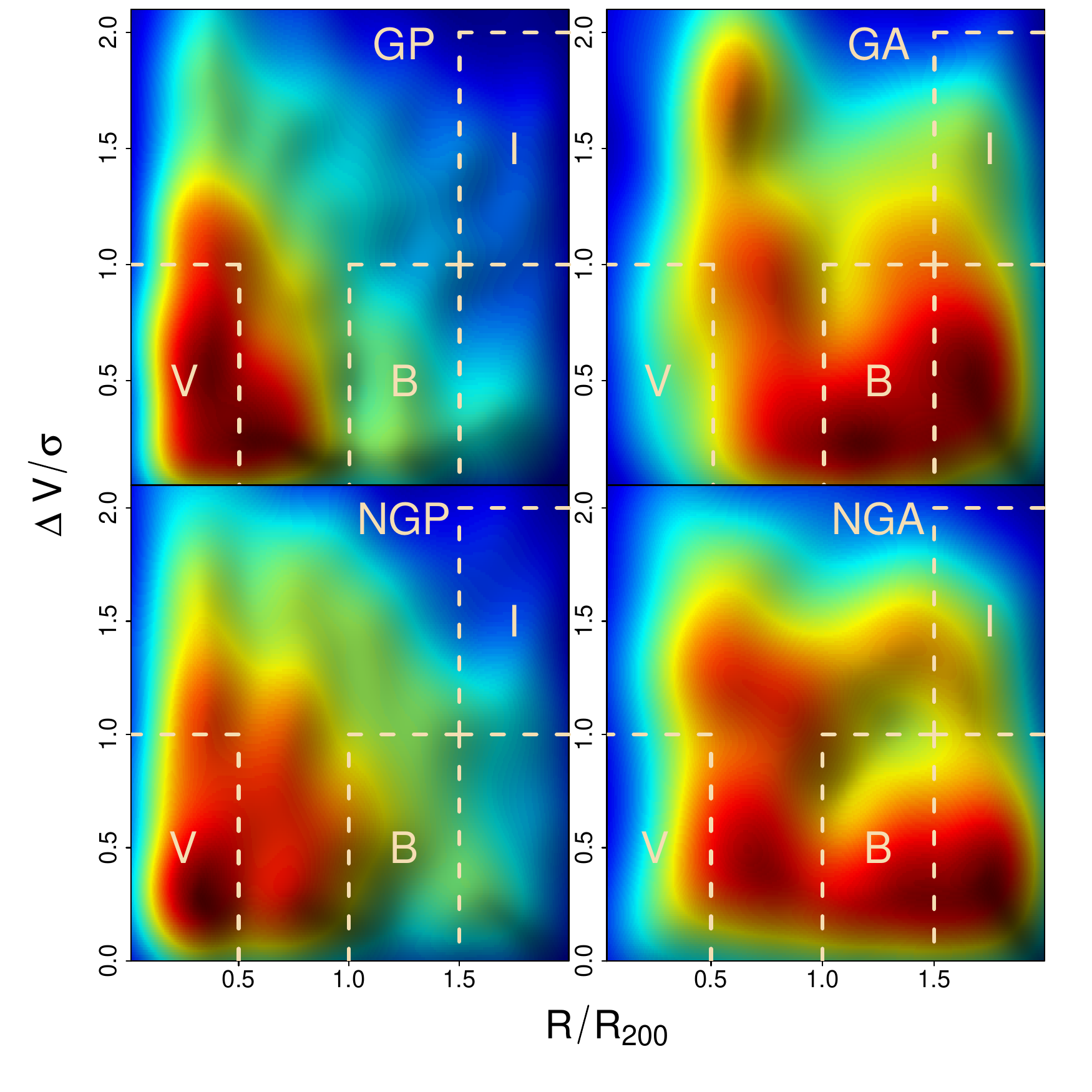} \\[\abovecaptionskip]
  %  \small (a) An image
  \end{tabular}

  %\vspace{\floatsep}
\vspace{-3.5cm}

  \begin{tabular}{@{}c@{}}
    \hspace{0.11cm}
    \includegraphics[scale=0.5]{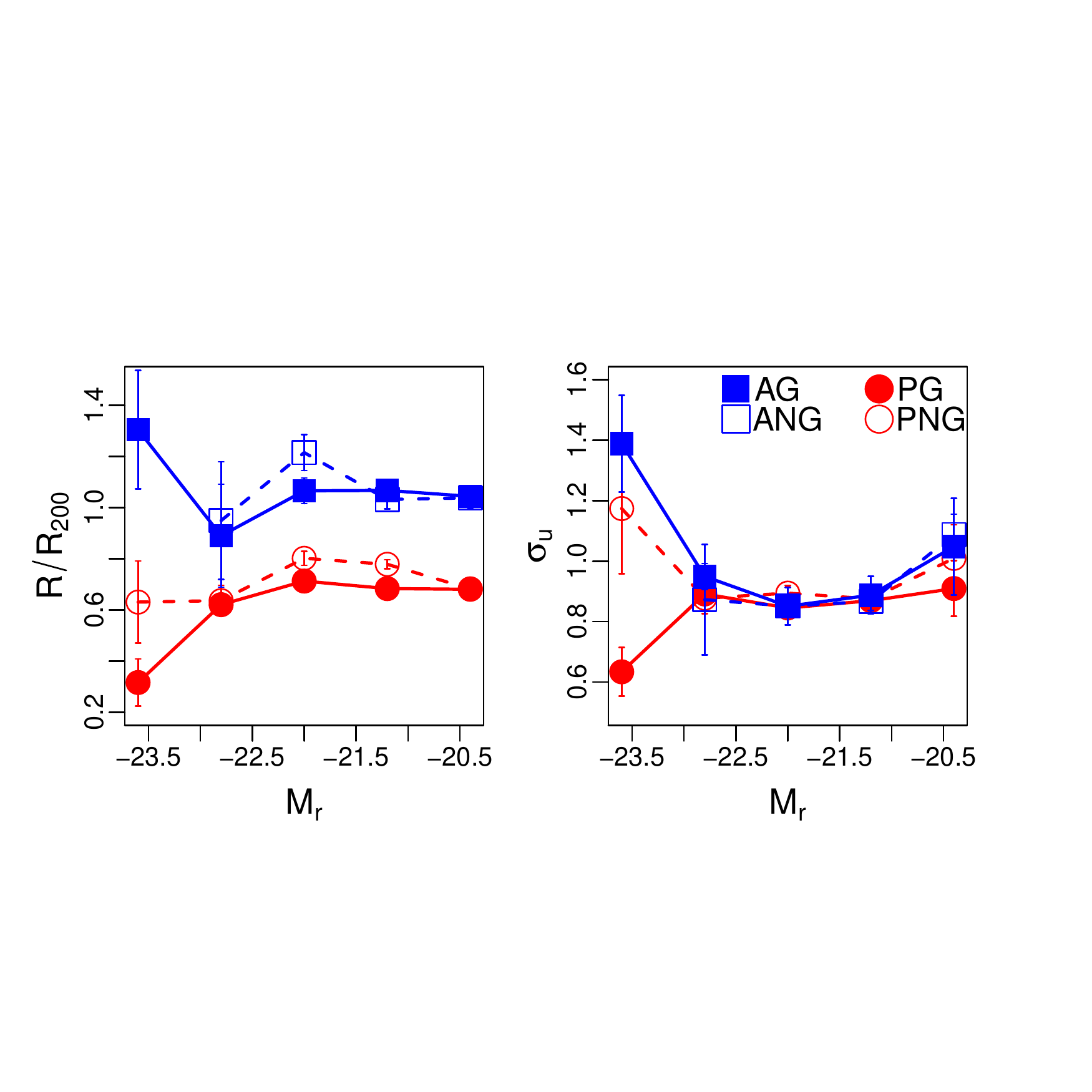} \\[\abovecaptionskip]
 %   \small (b) Another image
  \end{tabular}
\vspace{-2.8cm}
  \caption{Upper pannels: phase space diagrams for each sample. Redder colors indicate denser regions. Virial, backsplash, and infall regions are defined according to \citep{Mmr} and
 indicated by dashed lines. Clustercentric distance and velocity dispersion as a function of the absolute magnitude in the $r$ band for AG galaxies (filled blue squares), PG galaxies (filled red circles), ANG galaxies (open blue squares), and PNG galaxies (open red circles).}\label{fig:myfig}
\end{figure}

%\begin{figure}
%\begin{center}
%\hspace*{0.25cm}
%\vspace{5.5cm}
%\includegraphics[scale=0.5]{new_segreg_fig01.pdf}
%\vspace{-17.5cm}
%\hspace*{0.6cm}
%\end{figure}
%\vspace{-8cm}
%\begin{figure}
%\includegraphics[scale=0.4]{phase_space.pdf}
%\vspace{3.0cm}
%\caption{Clustercentric distance and velocity dispersion as a function of the absolute magnitude in the $r$ band for AG galaxies (filled blue circles), PG galaxies (filled red circles), ANG galaxies (open blue circles), and PNG galaxies (open red circles).}
%\label{fig2}
%\end{center}
%\end{figure}
\vspace{-0.5cm}

\section{ANALYSIS}
\label{Seg}

We start our analysis comparing the behaviour of the four families of galaxies defined in the previous section (PG, GA, PNG, ANG) in terms of velocity dispersion and projected clustercentric distance as a function of the absolute magnitude in the $r$ band. We also examine the distribution of galaxies in their respective phase-space diagrams. Next, we examine the velocity dispersion profiles (VDPs) of these families. Finally, we explore the behaviour of the VDPs with respect to the halo masses and galaxy luminosities.
\vspace{-0.4cm}

\subsection{Kinematic and luminosity segregation}

The upper panels of Figure 1 clearly show that the four samples occupy the phase-space in different ways. Note the higher density of passive galaxies (both PG and PNG) in smaller radii and especially within the the virial (V) region, with objects PG being even more concentrated than PNG. In comparison, note the higher density of AG and ANG objects in larger radii, especially within the backsplash (B) and infall (I) regions. This type of segregation suggests a connection between galaxy evolution and kinematics in the phase space.
The left lower panel of Figure 1 reinforces that PG have smaller clustercentric distances than the other families in all magnitude bins. The difference in relation to active galaxies, either AG or ANG, is clear in the plot. The difference in relation to PNG is less obvious, but applying the two-sample permutation test \citep{hig04}, we find that $\langle R/R_{200} \rangle_{\rm PG} = 0.7943 \pm 0.0095$ and $\langle R/R_{200} \rangle_{\rm PNG} =  0.8462 \pm 0.0128$, with p-value$=$0.0011, and thus we conclude that PG objects are more central than PNG at the 99\% c.l. Applying the permutation test on the AG and ANG samples we find $\langle R/R_{200} \rangle_{\rm GA} = 1.0611 \pm 0.0197$ and $\langle R/R_{200} \rangle_{\rm ANG} = 1.0577 \pm 0.0253$, with p-value$=$0.9164. Thus, these two families probably have the same radial distribution. Still in this plot, we see that the four families present different behaviour for $M_r \lesssim -22.75$. The PG and PNG samples have their brightest bins showing smaller clustercentric distances, but the effect is more pronounced for PG galaxies. 
%Removing the BCGs from the samples makes the effect less pronounced, but still significant.
On the other hand, the brightest bin of AG galaxies presents a greater value of radius, while the ANG sample presents a deficit of very luminous objects; see \citet{RLR} for a similar result. Finally, if we compare only passive and active objects, despite of being members of G or NG clusters, we find a strong spatial segregation, with passive galaxies having smaller clustercentric distances [$\langle R/R_{200} \rangle_{\rm P} = 0.8134 \pm 0.0008$] than active galaxies [$\langle R/R_{200} \rangle_{\rm A} = 1.0597 \pm 0.0015$], with p-value$~ < 10^{-4}$, which suggests a clear core-halo structure, irrespective of the dynamic state of the galaxy cluster.

Now, looking at the right lower panel of Figure 1, we see no important differences between the four families for $M_r \gtrsim -22.75$. This indicates a significant velocity mixing of galaxies in all the remaining magnitude bins. However, for values smaller than $M_r \simeq -22.75$ families split up dramatically: PG objects reach the lowest value of velocity dispersion, while AG and PNG show a signficant increase of velocity dispersion in comparison to the average behavior in the other magnitude bins. Once more, we see the deficit of bright ANG galaxies. These results indicate that only in G systems the passive and very luminous population slowed down toward the cluster center. Also, the brightest GA objects present high velocity dispersion and large radius, 
suggesting they are coming to the system via infall. Indeed, the fraction of bright GA objects ($M_r \gtrsim -21.5$) wrt all GA objects
in the I region (32\%) is a little higher than those we find in the
B (24\%) and V (25\%) regions. The fact that the brightest PNG objects have high velocity dispersion may be associated to the dynamic complexity of NG systems. 

Before proceeding to the next section,
it is important to say that tests with different data binning 
and/or removing the BCGs from each sample lead to non-significant changes of the results presented so far.
%\vspace{-0.4cm}
%All these findings indicate the importance of the orbit distribution of cluster galaxies, which brings us to our next topic of analysis: the velocity dispersion profiles of clusters.

\begin{figure}
\begin{center}
\hspace{-1.3cm}
\includegraphics[width=84mm]{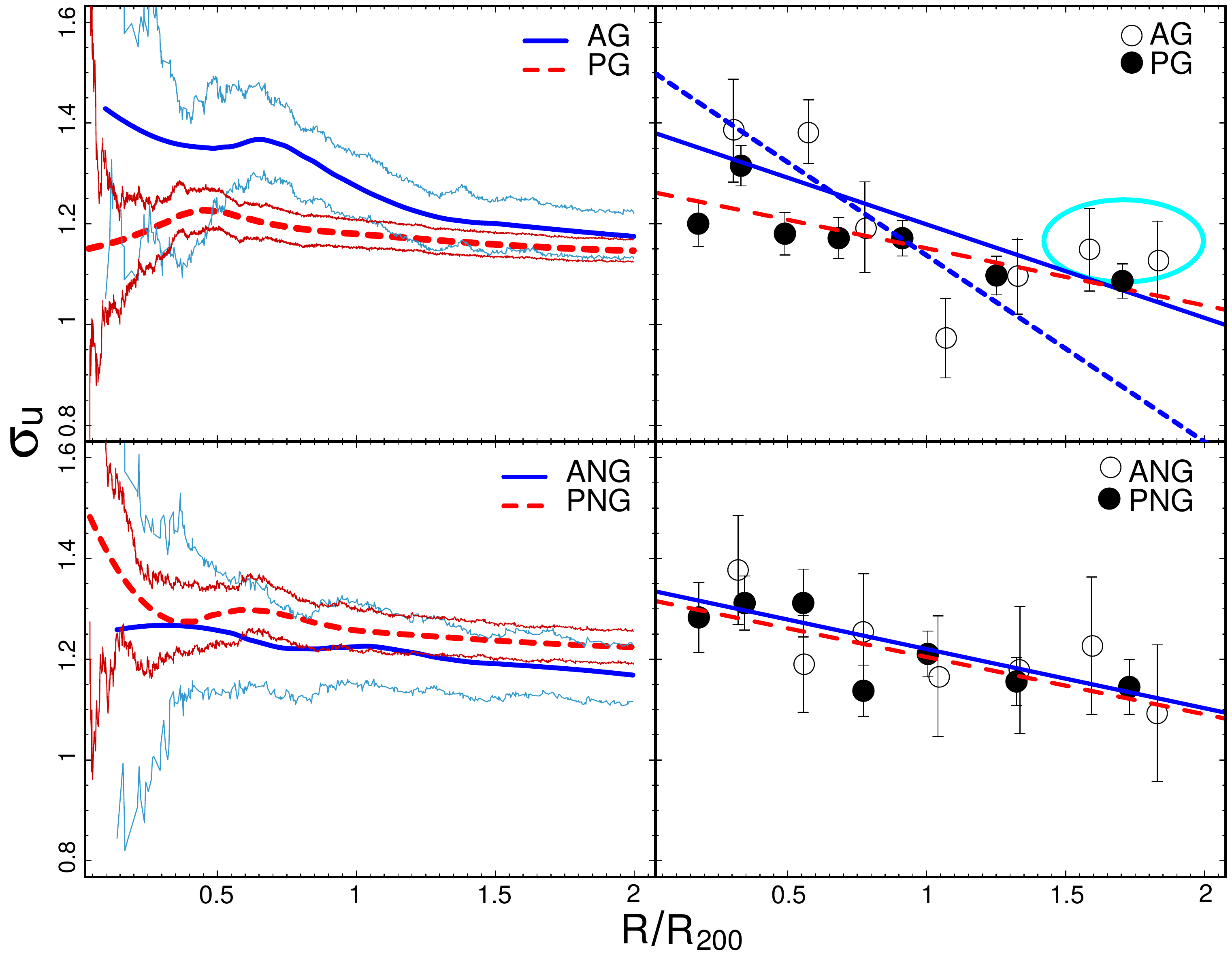}
\vspace{-0.35cm}
\caption{Left panels: cumulative VDPs for AG and ANG (blue lines) $+$ PG and PNG (red dashed lines) galaxies. Envelopes for 1,000 resamples are in red for PG an PNG galaxies, and in light blue for AG and ANG galaxies. Right panels: binned VDPs for AG and ANG (black open circles and blue lines) and PG and PNG galaxies (black filled circles and red dashed lines). A second VDP for AG objects, disregarding the last two points (marked in cyan), is depicted in dashed blue line.}
\end{center}
\end{figure}
\vspace{-0.1cm}

\subsection{Velocity dispersion profiles}

The study of VDPs has been used as a powerful tool for doing dynamical analysis of galaxy clusters \citep[see e.g.][]{stru,denK,pimb}. 
Recent studies indicate an important difference between VDPs with respect to the dynamical state of galaxy clusters \citep[see][]{H09,cava,CRD18,BP18}.
%The orbital distribution of the various types of galaxies in clusters can give us a unique information about the evolution of clusters themselves, and about the evolution of their member galaxies \citep{Bi04}. Recent studies also indicate an important difference between VDPs 
%with respect to the dynamical state of galaxy clusters \citep[see][]{H09,cava,CRD18}. 
In Figure 2 we plot the VDPs for all galaxy families in two versions, binned and cumulative. The cumulative VDPs were built following the procedure described in \citet{CRD18}. The use of both versions allows us to extract more information from the data. Firstly, note the marked differences between the cumulative profiles of the passive and active components. AG galaxies follow a decreasing profile, possibly a combination of isotropic and radial orbits \citep[see e.g.][]{Gi98,Bi04}, while PG galaxies present an almost flat profile, typical of objects with isotropic orbits -- except for the very central parts, where we see a small drop toward the center, indicating that circularization of orbits may be occurring in this component \citep{Gi98}. The corresponding binned VDPs reinforce this scenario. We use linear fits to measure how dominant are objects in radial orbits. We should expect that the isotropization of orbits should lead to slopes closer to 0 and intercepts closer to 1, and that passive galaxies have more isotropic orbits \citep[see e.g.][]{Bi04}. 
Analysis of covariance is used to test whether the slopes and intercepts are significantly different. In the upper right panel of Figure 2, the slope and intercept of the AG line are greater than those of the PG line. Using all AG points this occurs at the 90\% c.l., while  excluding the last two points the difference would be significant at the 95\% c.l. That is, the two components are dynamically distinct, but clusters are classified as G because the passive objects are dominant ($\sim$80\%) in these systems.
By contrast, in the lower right panel of Figure 2, the slopes and intercepts of both active and passive components are the same in NG clusters (p-value=0.9841), which supports the idea that these systems are dynamically similar. In the left lower panel of Figure 2, we see the passive component  presenting a slight central depression and then an increase from $\sim 0.5 - 1.0 R_{200}$. This subtle upward trend is also observed in NG clusters studied by \citet{H09} and \citet{CRD18}, and it can be interpreted as a signature of substructure or mergers \citep{MF96, cort}. Note that the AG VDP also presents an increase at similar radii. This can be interpreted as this component having some degree of perturbation that does not seem to be shared with the passive objects, and which should be insufficient to de-characterize the Gaussian velocity distribution. 
In the case of the PNG upturn, since this component is dominant ($\sim$79\%) this is probably the cause of non-gaussianity of these systems.

\vspace{-0.2cm}

\begin{figure}
\begin{center}
\hspace{-1.135cm}
\includegraphics[width = 84mm]{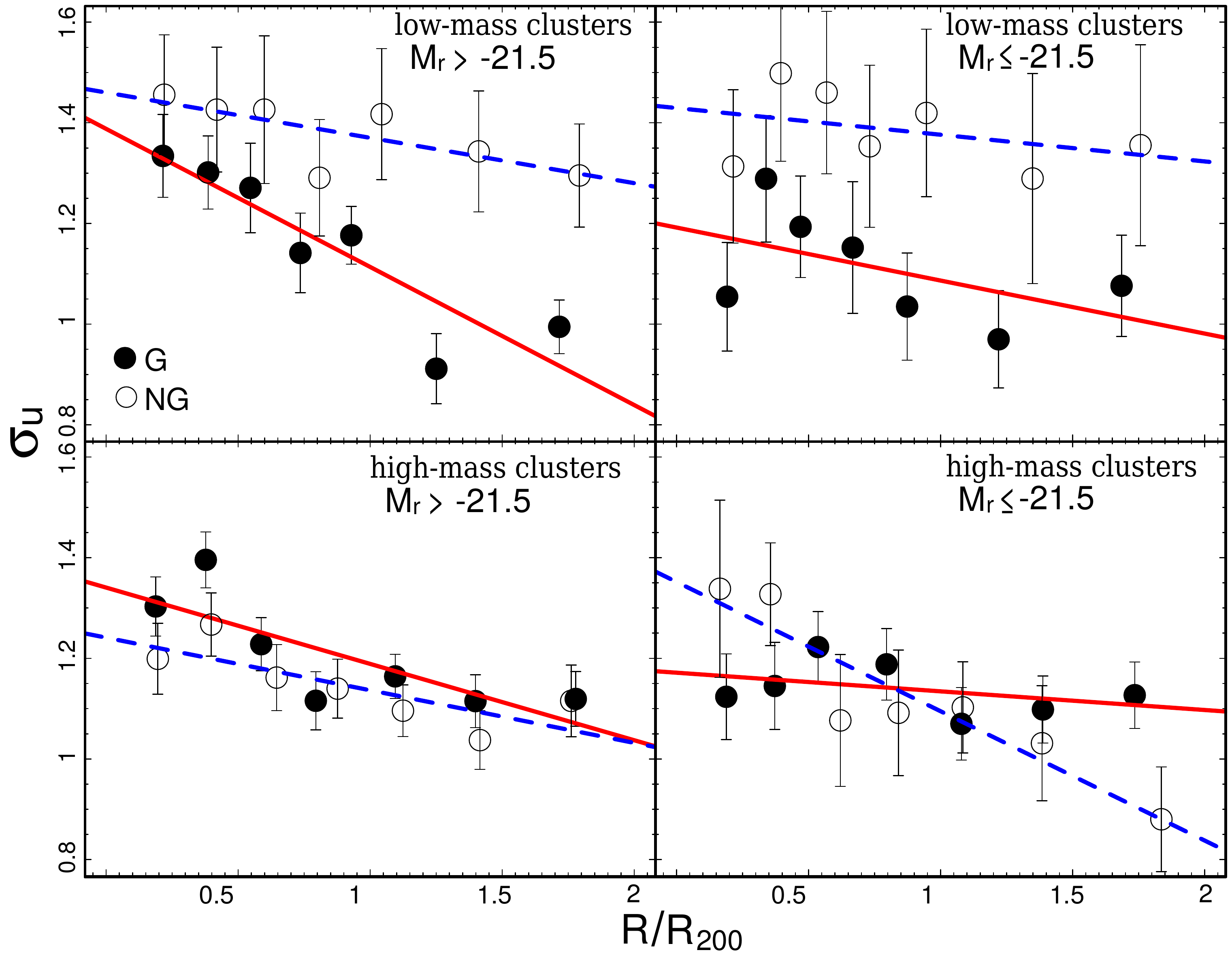}
\vspace{-0.4cm}
\caption{Binned VDPs for galaxies in G (filles circles and red lines) and NG systems (open circles and blue dashed lines). Upper and lower panel refer to rich and very rich clusters, respectively.
Left and right panels refer to galaxies with $M_r > -21.5$ and $M_r \leq -21.5$, respectively.}
\label{vdp}
\end{center}
\end{figure}
\vspace{-0.28cm}

\subsection{Halo masses and galaxy luminosities}

A further point we want to examine is the influence of the halo masses on the VDPs. Before doing this, we have to consider the problem of estimating the virial properties of NG systems. We use a correction based on iterative removal of galaxies whose absence in the sample makes clusters become Gaussian \citep{RLM11}. The corrected properties are just those the system would have if it were made only with galaxies consistent with the normal velocity distribution. This correction allows one to honestly compare typical properties of G and NG groups. After this procedure, we have clusters in the range $1.06 \times 10^{14} \text{M}_\odot \leq M_{200} \leq 1.55 \times 10^{15} \text{M}_\odot$, which were divided around the median value $\langle M_{200} \rangle = 3.43 \times 10^{14}\text{M}_\odot$, defining two subsamples: `low-mass clusters' ($\sim 57\%$ of the total sample) and `high-mass clusters' ($\sim 44\%$ of the total sample). To take into account the galaxy luminosities we have made a separation around the characteristic magnitude $M^\ast$ ($M_r=-21.5$; a separation around $M_r=-22.75$, as suggested in Section 3.1, would produce very unbalanced samples.). This corresponds to define two subsamples around this value with mean stellar masses  ${\rm \log{(M_\ast / M_{\sun})}=11.38}$ and ${\rm \log{(M_\ast / M_{\sun})}=10.66}$.  The following discussion takes into account significant differences between slopes and/or intercepts at the 95\% c.l. from comparing regressions via analysis of covariance.

In Figure 3, we present the respective binned VDPs.
The main point to highligh is that the slope and intercept of the linear fits for galaxies in G clusters decrease with both the halo masses and galaxy luminosities, indicating that brighter galaxies in richer clusters are possibly more dominated by objects in isotropic orbits. This trend does not appear in NG systems, for which we see different and more complex features. For rich and very rich NG systems, the less luminous galaxies have approximately the same slope (p-value=0.8719), with a significant decrease in the intercept of the linear fit for very rich clusters (p-value=0.0023). 
At the same time, the more luminous galaxies have more pronounced slopes in richer NG clusters (p-value=0.0477), but also present a significant decrease in the intercept (p-value=0.00401). A less clear evolutionary path in NG systems is consistent with the results presented in sections 3.1 and 3.2, probably indicating that neither galaxy family is completely virialized or sharing isotropic orbits. Still in Figure 3, note that the velocity segregation between G and NG systems is more pronounced for rich systems than for the very rich, which reinforces the importance of the halo mass on the evolutionary history of both clusters and their member galaxies.
\vspace{-0.5cm}

\section{CONCLUSIONS}

In this letter, we investigated the influence of the dynamic state of galaxy clusters on segregation phenomena and velocity dispersion profiles, trying to draw a link between these two types of cluster galaxies analyses. In Section 3.1, we found a strong spatial segregation between active (star forming) galaxies and passive ones both in G and NG systems. This segregation in projected radius indicates that clusters, despite its dynamic state, have a marked core-halo structure with the core being dominated by passive galaxies and the halo by active objects. We also found that the passive population in G systems is more concentrated than its counterpart in NG systems. This difference between passive galaxies located in G and NG clusters are reinforced by their very distinct VDPs, as showed in Section 3.2, with passive galaxies in G systems having more isotropic orbits. The similarity between the binned VDPs of ANG and PNG galaxies is a nice way to see how these populations, although distinct in their SFR, probably share a similar mix of orbits. Going back to section 3.1, we found that passive galaxies in G systems is the only family with lower velocity dispersions for the brightest galaxies ($M_r \lesssim -22.75$), thus presenting velocity segregation with luminosity. This result is complemented by the cumulative VDP in Figure 2, where we see a small but clear drop toward the center, suggesting these objects have slowed down toward the cluster centers, and showing that G systems are dynamically older. Complementing this picture, we see in Figure 3 a clear evolutionary trend for the VDPs of G systems, with brighter galaxies in richer clusters having flatter profiles, which may mean that they are more dominated by objects in isotropic orbits. The less straightforward picture for NG systems is probably due to a competition between their natural evolution  (violent relaxation + infall) which defines the core-halo pattern, and gravitational perturbations from mergers throughout their formation history, which can make the distribution of orbits more complex through these systems.
\vspace{-0.5cm}

\section*{Acknowledgments}
ALBR thanks the support of CNPq, grant 311932/2017-7. DFM thanks the financial support of CAPES.
PAAL thanks the support of CNPq, grant 308969/2014-6; and CAPES, process number 88881.120856/2016-01. RSN thanks the support of MCTIC, grant 301036/2017-9. We  also  would  like  to  thank  the  referee for  comments  and  suggestions  that  helped  improving  the manuscript.
\vspace{-0.5cm}



%\bibliographystyle{mnras}

\begin{thebibliography}{99}


\bibitem[\protect\citeauthoryear{Adami et al.}{1998}]{Ada98}Adami,C., Biviano, A., Mazure, A. 1998, A\&A, 331, 439

\bibitem[\protect\citeauthoryear{Amari}{1985}]{Am85} Amari, S., 1985, Differential-geometrical methods in statis- tics, Lecture Notes in Statistics, vol. 28, Springer

%\bibitem[\protect\citeauthoryear{Alpaslan et al.}{2016}]{Asp} Alpaslan, M., Grootes, M., Marcum, P.~M., et al.\ 2016, \mnras, 457, 2287 

%\bibitem[\protect\citeauthoryear{Balogh, Navarro \& Morris}{2000}]{Bl00} Balogh M. L., Navarro J. F.,Morris S. L., 2000, ApJ, 540, 113

%\bibitem[\protect\citeauthoryear{Balogh et al.}{2014}]{Ba14} Balogh, M.~L., McGee, S.~L., Mok, A., et al.\ 2014, \mnras, 443, 2679 

%\bibitem[\protect\citeauthoryear{Barnes \& Williams}{2012}]{BW12} Barnes, E.I, Williams, L.L.R., 2012, ApJ 748, 144

\bibitem[\protect\citeauthoryear{Barsanti et al. }{2018}]{bar18} Barsanti, S. et al. 2018, \apj, 857, 71

\bibitem[Beck et al.(2016)]{Bk16} Beck, R., Dobos, L., Budav{\'a}ri, T., et al.\ 2016, \mnras, 460, 1371.

%\bibitem[\protect\citeauthoryear{Beraldo e Silva et al.}{2013}]{Be13} Beraldo e Silva, L.~J., Lima, M., \& Sodr{\'e}, L.\ 2013, \mnras, 436, 2616 

\bibitem[\protect\citeauthoryear{Berlind et al.}{2006}]{Ber06} Berlind, A.~A., Frieman, J., Weinberg, D.~H., et al.\ 2006, ApJS, 167, 1 

\bibitem[\protect\citeauthoryear{Bilton \& Pimbblet}{2018}]{BP18} Bilton L.E. and Pimbblet K.A., MNRAS 2018, 481, 1507

\bibitem[\protect\citeauthoryear{Biviano et al.}{1992}]{Bi92} Biviano, A., Girardi, M., Giuricin, G., Mardirossian, F., Mezzetti, M.\ 1992, \apj, 396, 35 

\bibitem[\protect\citeauthoryear{Biviano \& Katgert}{2004}]{Bi04} Biviano, A. \& Katgert, P. 2004, \aap,
424, 779

\bibitem[\protect\citeauthoryear{Biviano et al.}{2006}]{Bi06} Biviano, A., Murante, G., Borgani, S., et al. 2006, \aap, 456, 23

%\bibitem[\protect\citeauthoryear{Boselli \& Gavazzi}{2006}]{Boga} Boselli, A. \& Gavazzi, G. 2006, PASP, 118, 517

\bibitem[\protect\citeauthoryear{Brinchmann et al.}{2004}]{B04} Brinchmann, J., Charlot, S., White, S. D. M., et al. 2004, MNRAS, 351, 1151

%\bibitem[\protect\citeauthoryear{Brough}{2006}]{br06} Brough S., Forbes D.A., Kilborn V.A., Couch W., 2006, MNRAS, 370, 1223
\bibitem[\protect\citeauthoryear{Carlberg et al.}{1997}]{carl97} Carlberg, R.G. et al. 1997, \apj, 476, L7

\bibitem[\protect\citeauthoryear{Cava et al.}{2017}]{cava} Cava, A. et al. 2017, \aap, 606, id.A108

%\bibitem[\protect\citeauthoryear{Contini \& Kang}{2015}]{kc16} Contini, E., \& Kang, X.\ 2015, \mnras, 453, L53 

\bibitem[\protect\citeauthoryear{Cortese et al.}{2004}]{cort} Cortese, L., Gavazzi, G., Boselli, A., Iglesias-Paramo, J. \& Carrasco, L. 2004, \aap, 425, 429


\bibitem[\protect\citeauthoryear{Costa et al.}{2018}]{CRD18} Costa, A. P., Ribeiro, A.~L.~B., \& de Carvalho, R.~R., 2018, MNRAS, 473, L31

\bibitem[\protect\citeauthoryear{de Carvalho et al.}{2017}]{deC17} de Carvalho, R.~R., Ribeiro, A.~L.~B., Stalder, D., et al.\ 2017, \aj, 154, 96 

\bibitem[\protect\citeauthoryear{den Hartog \& Katgert}{1996}]{denK} den Hartog R., Katgert P., 1996, \mnras, 279, 349

%\bibitem[\protect\citeauthoryear{De Propris et al.}{2014}]{DeP} De Propris, R., Baldry, I.~K., Bland-Hawthorn, J., et al.\ 2014, \mnras, 444, 2200 

%\bibitem[\protect\citeauthoryear{Dressler \& Gunn}{1983}]{DG83} Dressler A., Gunn J.E. 1983, ApJ, 270, 7

%\bibitem[\protect\citeauthoryear{Dressler \& Schectman}{1988}]{DS88} Dressler A. \& Schectman, S. A., 1988, AJ, 95, 985

%\bibitem[\protect\citeauthoryear{Fraley \& Raftery}{2007}]{FR07} Fraley, C \& Raftery, A.E., 2007, Journal of Classification, 24,155

%\bibitem[\protect\citeauthoryear{Gill et al.}{2005}]{GKG} Gill S. P. D., Knebe A., Gibson B. K., 2005, MNRAS, 356, 1327

\bibitem[\protect\citeauthoryear{Girardi et al.}{1998}]{Gi98} Girardi, M., Giuricin, G., Mardirossian, F., et al. 1998, ApJ, 505, 74

\bibitem[\protect\citeauthoryear{Girardi et al.}{2003}]{Gi03} Girardi,  M.,  Rigoni,  E.,  Mardirossian,  F.,  Mezzetti,  M.  2003, \aap, 406, 403

%\bibitem[\protect\citeauthoryear{Goto et al.}{2003}]{go3} Goto, T., Okamura, S., Sekiguchi, M., et al.\ 2003, \pasj, 55, 757 

\bibitem[\protect\citeauthoryear{Goto}{2005}]{go1} Goto, T. 2005, MNRAS, 359, 1415

\bibitem[\protect\citeauthoryear{Higgins}{2004}]{hig04} Higgins, J. J. (2004) Introduction to Modern Nonparametric Statistics.


\bibitem[\protect\citeauthoryear{Hou et al.}{2009}]{H09} Hou, A., Parker, L.~C., Harris, W.~E., Wilman, D.~J.\ 2009, \apj, 702, 1199 

\bibitem[\protect\citeauthoryear{Hou et al.}{2012}]{H12} Hou, A., Parker, L.~C., Wilman, D.~J., et al.\ 2012, \mnras, 421, 3594 

%\bibitem[\protect\citeauthoryear{Jarque \& Bera}{1980}]{JB80} Jarque, C. M. \& Bera, A. K., 1987, International Statistical Review 55, 163

%\bibitem[\protect\citeauthoryear{Joshi et al.}{2016}]{jo16} Joshi, G.~D., Parker, L.~C., \& Wadsley, J.\ 2016, \mnras, 462, 761 

%\bibitem[\protect\citeauthoryear{Kauffmann et al.}{2003}]{k03} Kauffmann, G., Heckman, T. M., White, S. D. M., et al. 2003, MNRAS, 341, 33 

%\bibitem[\protect\citeauthoryear{Krause et al.}{2013}]{KRL13} Krause, M.~O., Ribeiro, A.~L.~B., \& Lopes, P.~A.~A.\ 2013, \aap, 551, A143 

\bibitem[\protect\citeauthoryear{La Barbera et al.}{2010}]{Lab10} La Barbera, F., Lopes, P.~A.~A., de Carvalho, R.~R., de La Rosa, I.~G., Berlind, A.~A.\ 2010, MNRAS, 408, 1361

\bibitem[\protect\citeauthoryear{Lares et al.}{2004}]{la} Lares, M., lambas, D. G., S\'anchez, A. G., 2004, MNRAS, 352, 501

%\bibitem[\protect\citeauthoryear{Lin et al.}{2010}]{Lin10} Lin, L., Cooper, M.~C., Jian, H.-Y., et al.\ 2010, \apj, 718, 1158 

%\bibitem[\protect\citeauthoryear{Lynden-Bell}{1967}]{ly67} Lynden-Bell, D., 1967, MNRAS, 136, 101

\bibitem[\protect\citeauthoryear{Lopes}{2007}]{Lop07}Lopes, P. A. A. 2007, MNRAS, 380, 1608

\bibitem[\protect\citeauthoryear{Lopes et al.}{2009}]{Lop09}Lopes, P. A. A., de Carvalho, R. R., Kohl-Moreira, J. L., et al. 2009, MNRAS, 392, 135

%\bibitem[\protect\citeauthoryear{Ma et al.}{2010}]{Ma10} Ma, C.-J., Ebeling, H., Marshall, P., \& Schrabback, T. 2010, MNRAS, 406, 121

\bibitem[\protect\citeauthoryear{Mahajan et al.}{2011}]{Mmr} Mahajan, S., Mamon, G. A, \& Raychaudhury, S. 2011, MNRAS, 416, 2882

\bibitem[\protect\citeauthoryear{Mart{\'{\i}}nez \& Zandivarez}{2012}]{Mz12} Mart{\'{\i}}nez, H.~J., \& Zandivarez, A.\ 2012, \mnras, 419, L24 

\bibitem[\protect\citeauthoryear{Menci \& Fusco-Femiano}{1996}]{MF96} Menci, N., \& Fusco-Femiano, R.\ 1996, \apj, 472, 46 

%\bibitem[\protect\citeauthoryear{Merrall \& Henriksen}{2003}]{MH03} Merrall, T.E.C., Henriksen, R.N., 2003, \apj, 595, 43

%\bibitem[\protect\citeauthoryear{McIntosh et al.}{2005}]{McI} McIntosh, D.~H., Zabludoff, A.~I., Rix, H.-W., \& Caldwell, N.\ 2005, \apj, 619, 193 
\bibitem[\protect\citeauthoryear{Moss \& Dickens}{1977}]{MD} Moss, C. \& Dickens, R.J. 1977, MNRAS, 178, 701
\bibitem[\protect\citeauthoryear{Nascimento, Ribeiro \& Lopes}{2017}]{NRL}Nascimento, R.S., Ribeiro, A.L.B., \& Lopes, P.A.A., 2017, \mnras, 464, 183
\bibitem[\protect\citeauthoryear{Old et al.}{2015}]{old15}
Old, L., Wojtak, R., Mamon, G. A., et al., 2015, \mnras, 449, 1897
%\bibitem[\protect\citeauthoryear{Natarajan et al.}{1997}]{NHV97} Natarajan, P., Hjorth, J., \& van Kampen, E.\ 1997, \mnras, 286, 329 

\bibitem[\protect\citeauthoryear{Oman \& Hudson}{2016}]{OH16} Oman, K.A. \& Hudson, M.J.
2016, \mnras, 463, 3083

\bibitem[\protect\citeauthoryear{Pimbblet et al.}{2014}]{pimb} Pimbblet K. A., Penny S. J., Davies R. L., 2014, \mnras, 438, 3049

\bibitem[\protect\citeauthoryear{Popesso et al.}{2005}]{Pop05} Popesso, P., Biviano, A., Bohringer, H., Romaniello, M., Voges, W., 2005, A\&A, 433, 431 

\bibitem[\protect\citeauthoryear{Popesso et al.}{2007}]{Pop07} Popesso, P., Biviano, A., Bohringer, H., Romaniello, M., 2007, A\&A, 464, 451 

%\bibitem[\protect\citeauthoryear{Poudel et al.}{2016}]{Pou} Poudel, A., Hein{\"a}m{\"a}ki, P., Nurmi, P., et al.\ 2016, \aap, 590, A29 

%\bibitem[\protect\citeauthoryear{Presotto et al.}{2012}]{Pre12} Presotto, V., Iovino, A., Scodeggio, M., et al.\ 2012, \aap, 539, A55

%\bibitem[\protect\citeauthoryear{Ribeiro et al.}{2009}]{R09} Ribeiro, A.~L.~B., Trevisan, M., Lopes, P.~A.~A., \& Schilling, A.~C.\ 2009, \aap, 505, 521 

\bibitem[\protect\citeauthoryear{Ribeiro et al.}{2010}]{RLM10} Ribeiro, A.~L.~B., Lopes, P.~A.~A., Trevisan, M.\ 2010, \mnras, 409, L124 

\bibitem[\protect\citeauthoryear{Ribeiro et al.}{2011}]{RLM11} Ribeiro, A.~L.~B., Lopes, P.~A.~A., Trevisan, M.\ 2011, \mnras, 413, L81 

\bibitem[\protect\citeauthoryear{Ribeiro et al.}{2013a}]{R13} Ribeiro, A.~L.~B., de Carvalho, R.~R., Trevisan, M., et al.\ 2013, \mnras, 434, 784 

\bibitem[\protect\citeauthoryear{Ribeiro et al.}{2013b}]{RLR} Ribeiro, A.~L.~B., Lopes, P.~A.~A., \& Rembold, S.~B.\ 2013, \aap, 556, A74 

%\bibitem[\protect\citeauthoryear{Roberts et al.}{2015}]{R15} Roberts, I.~D., Parker, L.~C., Joshi, G.~D., \& Evans, F.~A.\ 2015, \mnras, 448, L1 

\bibitem[\protect\citeauthoryear{Roberts et al.}{2018}]{rob18} Roberts, I.~D., Parker, L.~C. \& Hlavacek-Larrondo, J. 2018, \mnras, 475, 4704


%\bibitem[\protect\citeauthoryear{Robotham et al.}{2014}]{Rob} Robotham, A.~S.~G., Driver, S.~P., Davies, L.~J.~M., et al.\ 2014, \mnras, 444, 3986 

\bibitem[\protect\citeauthoryear{Ruckdeschel}{2006}]{ruck} Ruckdeschel, P. 2006, Metrika, 63, 295

\bibitem[\protect\citeauthoryear{Sodr\'e et al.}{1989}]{Scs89} Sodr\'e, L., Jr., Capelato, H.~V., Steiner, J.~E., \& Mazure, A.\ 1989, \aj, 97, 1279 

\bibitem[\protect\citeauthoryear{Salim et al.}{2007}]{S07} Salim, S., Rich, R. M., Charlot, S., et al. 2007, ApJS, 173, 267

%\bibitem[\protect\citeauthoryear{Strateva}{2001}]{st01} Strateva I., Ivezi\'c Z., Knapp G. et al. 2001, AJ, 122, 1861

\bibitem[\protect\citeauthoryear{Struble}{1979}]{stru} Struble M. F., 1979, \aj, 84, 27

\bibitem[\protect\citeauthoryear{Stein}{1997}]{st97} Stein, P., 1997, A\&A, 317, 670

%\bibitem[\protect\citeauthoryear{Taranu et al.}{2014}]{Ta14} Taranu, D.~S., Hudson, M.~J., Balogh, M.~L., et al.\ 2014, \mnras, 440, 1934 

%\bibitem[\protect\citeauthoryear{van den Bosh et al.}{2008}]{van08} van den Bosch, F.~C., Aquino, D., Yang, X., et al.\ 2008, \mnras, 387, 79 

%\bibitem[\protect\citeauthoryear{von der Linden  et al.}{2010}]{von10} von der Linden A.,Wild V., Kauffmann G.,White S.D.M., Weinmann S., 2010, MNRAS, 404, 1231

%\bibitem[\protect\citeauthoryear{Vulcani et al.}{2013}]{vu13} Vulcani, B., Poggianti, B.~M., Oemler, A., et al.\ 2013, \aap, 550, A58 

%\bibitem[\protect\citeauthoryear{Weinmann et al.}{2006}]{We06} Weinmann, S.~M., van den Bosch, F.~C., Yang, X., \& Mo, H.~J.\ 2006, \mnras, 366, 2 

\bibitem[\protect\citeauthoryear{Wetzel et al.}{2012}]{wtc12} Wetzel, A.~R., Tinker, J.~L., \& Conroy, C.\ 2012, \mnras, 424, 232 

\bibitem[\protect\citeauthoryear{Yahil \& Vidal}{1977}]{YV77} Yahil A. \& Vidal N. V., 1977, ApJ, 214, 347

\bibitem[\protect\citeauthoryear{Yang et al.}{2007}]{y07} Yang, X. et al. 2007, \apj, 671, 153

\bibitem[\protect\citeauthoryear{Yee \& L\'opes-Cruz}{1999}]{YL99} Yee H. \& L\'opez-Cruz O., 1999, AJ, 117, 1985

\bibitem[\protect\citeauthoryear{Zabludoff \& Franx}{1993}]{zbf} Zabludoff, A. \& Franx, M., \aj, 106, 	1314

%\bibitem[\protect\citeauthoryear{Zabludoff et al.}{1996}]{zbd} Zabludoff A.I. et al. 1996, ApJ, 466, 104

%\bibitem[\protect\citeauthoryear{Zabludoff \& Mulchaey}{1998}]{zpb} Zabludoff A.I. \& Mulchaey J.S., 1998, \apj, 496, 39

%\bibitem[\protect\citeauthoryear{Ziparo et al.}{2013}]{zi13} Ziparo, F., Popesso, P., Biviano, A., et al.\ 2013, \mnras, 434, 3089 

\end{thebibliography}


\end{document}